\documentclass[preprint,12pt]{elsarticle}
\usepackage{amsmath, amsfonts, amssymb, amsthm, euscript, amscd, latexsym, mathrsfs, bm, xcolor,accents, setspace}
\usepackage{bbm}
\usepackage{enumitem}
\usepackage{lipsum}
\usepackage{hyperref}
\def\F{{\mathcal F}}

\def\NN{\mathbb{N}}

\def\RR{\mathbb{R}}

\newtheorem{Thm}{Theorem}

\newtheorem{Lem}[Thm]{Lemma}
\newtheorem{Cor}[Thm]{Corollary}

\newtheorem{Rmk}{Remark}

\def\aa#1{ \begin{align*} #1 \end{align*} }
\def\aaa#1{ \begin{align} #1 \end{align} }
\def\mm#1{ \begin{multline*} #1 \end{multline*} }
\def\mmm#1{ \begin{multline} #1 \end{multline} }
\newcommand{\lt}{\leqslant}
\newcommand{\gt}{\geqslant}
\def\si{\sigma}
\def\al{\alpha}

\def\dl{\delta}

\def\td{\tilde}
\def\eps{\varepsilon}
\def\begeq{\begin{equation} \begin{cases}} 
\def\endeq{ \end{cases} \end{equation}}
\def\eq#1{ \begeq #1 \endeq }

\newcommand{\E}{{\mathbb E}}

\newcommand{\x}{\times}

\newcommand{\la}{\lambda}

\newcommand{\fdot}{\,\cdot\,}

\newcommand{\om}{\omega}

\newcommand{\rf}{\eqref}
\newcommand{\lb}{\label}

\newcommand{\mto}{\mapsto}

\newcommand{\diag}{{\rm diag}}

\newcommand{\mc}{\mathcal}

\newcommand{\sg}{\sigma}
\renewcommand{\leq}{\lt}

\renewcommand{\>}{\big\rangle}

\newcommand{\Dl}{\Delta}
\newcommand{\hsp}{{\hspace{-2mm}}}
\newcommand{\hs}{{\hspace{-1mm}}}

\newcommand{\bi}{\begin{itemize}}
\newcommand{\ei}{\end{itemize}}

\begin{document}

\begin{frontmatter}
\title
{Hedging in a market with jumps  -- an FBSDE approach}

%\lb{largeinvestorlevy}

\author{Rui S\'a Pereira}
\address{Departamento de Matem\'atica, Universidade do Porto, 
%Rua Campo Alegre 687, 
Porto, Portugal}
\ead{manuelsapereira@gmail.com}

\author{Evelina Shamarova}
\address{Departamento de Matem\'atica, Universidade Federal da Para\'iba, Jo\~ao Pessoa, Brazil}
\ead{evelina@mat.ufpb.br}

%\maketitle

\begin{abstract}
We propose a model for hedging  in a market with jumps for a large investor. 
The dynamics of the stock prices and the value process %of the hedger 
is governed by forward-backward SDEs driven by Teugels martingales. 
 Unlike known FBSDE market models, ours accounts for jumps in stock prices.  
 Moreover, it allows to find an optimal hedging strategy. 
\end{abstract}


\end{frontmatter}

\section{Introduction}

In this article, 
we propose a forward-backward-SDE (FBSDE) model for hedging  in a market with a large investor and where the stock prices are allowed to jump.
 The attempts to price contingent claims
  have their origin in the work of  \cite{bs}, where the authors give a formula to price a european call option %under no-arbitrage conditions, 
  assuming that the stock price evolves as a geometric Brownian motion.
 One of the assumptions of the classical Black and Scholes model is that no individual investor  action is able to influence market prices.   The importance of accounting for the existence of large investors, however, has been increasing, given the prevalence of electronic trading and, in particular, high frequency trading which makes possible to issue thousands of orders over short periods of time.
   This problem was addressed by \cite{cvitanicma}, where the authors 
   present an FBSDE market model for a large investor, but in
   a Brownian market environment, where the stock prices, being modeled as geometric Brownian motions, 
   are not allowed to have jumps.
However,  the documented evidence  of  jumps in the distribution of the returns (see, e.g.,  \cite{eberlein}) suggests that
   a geometric Brownian motion  is not entirely suited to model the evolution of stock prices in real markets. 
   In particular, in periods of heavy market turbulence, such as the ``flash crash'' in May 2010, when the main US indexes temporarily dropped by more than 9 per cent,  hedging strategies resulting from Brownian models  leave investors exposed to a significant downside risk. 

It is known that markets, where the stock prices are modeled involving L\'evy processes are, in general, incomplete, 
so contingent claims may not admit self-financing replicating strategies. 
The first attempt to define optimal strategies in the context of incomplete markets  was made by 
 \cite{folm-schw}, where the authors  propose an optimal strategy as the one that minimizes, in a certain sense,  the injection of capital needed.

In our FBSDE market model, the evolution of the 
$d$-dimensional stock price $S_t = \{S_t^i\}_{i=1}^d$
 is governed by an SDE driven by  
$m$ independent Brownian motions and $d-m$ martingales 
picked from the system of orthonormalized Teugels martingales
$\{H^{(ik)}_t\}_{i=1, \, k\in \NN}^l$ such that %$i=1, \ldots, l$, with even indexes $k\in \NN$.
%(see \cite{nualart} for the definition),
for each $i\in \{1, \dots, l\}$, the family $\{H^{(ik)}_t\}_{k=1}^\infty$ is associated to a L\'evy process $L^i_t$.
All processes $L^i_t$ are assumed  independent and purely discontinuous.
 We refer the reader to \cite{nualart} (p. 763) for details on the martingales $H^{(ik)}_t$.
 Remark that different stock prices $S^i_t$ can jump at different times.
 Further, the value process $V_t$ and the portfolio process $\pi_t = \{\pi^i_t\}_{i=1}^d$ 
evolve according to a backward SDE with the final condition 
$h(S_T)$ which is the payoff at maturity $T$. 

Our model involves the martingales $H^{(ik)}_t$ because they are independent, strongly orthonormal,  purely discontinuous,
but most importantly, the system $\{H^{(ik)}_t\}_{i=1, \, k\in \NN}^l$, completed with the Brownian motions $\{B^i_t\}_{i=1}^{m}$,
possesses the predictable representation property. The latter allows 
to decompose the discounted value process into a sum of the value of the hedging portfolio and a strongly orthogonal martingale.
Therefore, our model allows to find a hedging strategy which is optimal in the sense of \cite{schweizer}. 
%because they share some properties with Brownian motions 
%(e.g., they are independent, strongly orthonormal, have variance $t$),
%but also, they are purely discontinuous. 
It is worth to mention that due to the presence of $H^{(ik)}_t$'s,  the SDEs representing the evolution of stocks become, in fact, driven by power-jump martingales
built on the basis of the underlying L\'evy processes (see \cite{nualart}, p. 763). The presence of these ``power-jump'' terms may
reflect ``skewness'', ``kurtosis'', and other volatile behavior or extremal movements of the market.
%Importantly, the predictable representation property of the system
%$\{H^{(ik)}_t\}_{i=1, \, k\in \NN}^l$, completed with the Brownian motions $\{B^i_t\}_{i=1}^{m}$,
%allows to decompose the discounted value process into a sum of the value of the hedging portfolio and a strongly orthogonal martingale.
%Therefore, our model allows to find a hedging strategy which is optimal in the sense of \cite{schweizer}. 

Thus, the main contribution of this work is introducing a model which accounts for jumps in stock prices and allows to find an optimal hedging strategy 
in the context of incomplete markets. %with jumps.

\section{FBSDE model for hedging in a market with jumps}

%We fix $T>0$ and 
In what follows, we present our model. Let  $(\Omega, {\mathcal F}, P)$  be a 
probability space, $\{B^i_t\}_{i=1}^m$ be independent real-valued  Brownian motions,
and $\{L^i_t\}_{i=1}^l$ be independent purely discontinuous  real-valued L\'evy processes with the L\'evy measures $\nu^i$
satisfying 
\aaa{
\lb{a1}
\int_{\RR}(1 \wedge x^2) \nu^i(dx) < \infty \quad \text{and} \quad  \int_{|x| \gt \eps} e^{\la|x|} \nu^i(dx) < C
}
 for some positive $\eps,\la$, and $C$.  
 Define the filtration  $ \mc F_t =  \sg\{B_s^i, 0 \lt s\lt t, \:  1 \lt i  \lt m \} \vee \sg\{L^i_s , 0 \lt s \lt t, \: 1 \lt i \lt l\}  \vee \mc N$, 
 where $\mc N$ is the collection of all $P$-null sets. We agree that all L\'evy measures $\nu$ considered in this work, satisfy the condition $\nu(\{0\})=0$.
 
 Let,  for each $i\in \{1,\ldots, l\}$, $\{H^{(ik)}_t\}_{k=1}^\infty$   
 be the family of orthonormalized Teugels martingales associated to the L\'evy process $L^i_t$.
 Lemma \ref{hi-rep} below provides 
 a useful representation for the orthonormalized Teugels martingales $H^{(i)}_t$ associated 
to an arbitrary one-dimensional L\'evy process $\ell_t$ 
with the L\'evy triple $(b,a,\la)$, $a = (a_1, \ldots, a_M)$, and the  L\'evy measure $\la$
 satisfying \rf{a1} with $\nu_i$ substituted by $\la$.
As in \cite{nualart}, we introduce the polynomials $q_{i-1}(x)$ obtained by the orthonormalization
of the system $\{1, x, x^2, \ldots\}$ with respect to the measure $x^2\la(dx) + |a|^2 \dl_0(dx)$,
where $\dl_0$ is the Dirac measure. Furthermore,  we define $p_i(x) = x q_{i-1}(x)$.
\begin{Lem}
\lb{hi-rep} 
Let $\ell$ be a one-dimensional L\'evy process with the 
L\'evy-It\^o decomposition 
$\ell_t = bt +  \sum_{i=1}^M a_i \beta_i (t) + \int_{|x| \lt 1 } x \td \mu(t,dx) + \int_{|x| > 1} x \mu(t,dx)$,
where  
 $\{\beta_i(t)\}_{i=1}^M$ are independent real-valued standard Brownian motions. 
 Then, it holds that %$H^{(i)}_t$ has the representation
 $H_t^{(i)}= q_{i-1}(0)  \sum\nolimits_{j=1}^M a_j \beta_j (t)+ \int_\RR p_i(x)\tilde{\mu}(t,dx)$.
In particular, if $\ell_t$ is purely discontinuous, then
$H_t^{(i)} = \int_\RR p_i(x)\tilde{\mu}(t,dx)$.
 \end{Lem}
\begin{proof}
Define $\td p_i(x) = p_i(x) - xq_{i-1}(0)$.
We will use the following representation for $H^{(i)}_t$ obtained in
\cite{nualart} (p. 763): \\
$H_t^{(i)} = q_{i-1}(0)\ell_t + \sum_{0 < s \leq t}
\tilde{p}_i(\Delta \ell_s) - t\, \E \bigg[ \sum_{0 < s \leq 1}
\tilde{p}_i(\Delta \ell_s)\bigg] - tq_{i-1}(0)\E[\ell_1].$\\
 Since $\ell_t= \ell_t^c + \sum_{0 \lt s \lt t} \Delta \ell_s$, where $\ell^c_t$
 is the continuous part of $\ell_t$, we obtain \\
 $H_t^{(i)}  = q_{i-1}(0)\ell_t^c + \sum_{0 < s \lt t}p_i(\Delta \ell_s)
 - \E\big[\sum_{0 < s \lt t} \tilde{p}_i (\Delta \ell_s)\big] -q_{i-1}(0)\E[\ell_t]$\\
\phantom{$H_t^{(i)}$} $= q_{i-1}(0)\big[\ell_t^c - \E[\ell_t^c]\big] + \sum_{0 < s \lt t}p_i(\Delta \ell_s) - \E\big[ \sum_{0 < s \lt t}p_i(\Delta \ell_s)\big]$\\
\phantom{$H_t^{(i)}$} $= q_{i-1}(0) \sum_{j=1}^M a_j \beta_j (t) +  \int_\RR p_i(x) \tilde{\mu} (t, dx)$.
\end{proof}

 Let us proceed with the description of the model. Fix a finite-time horizon $T>0$ and consider a market consisting of
$d$ risky assets (stocks) and risk-free money on a deposit. 
We assume that the price process of the risk-free deposit evolves according to the equation
\aaa{
\lb{bond}
dD_t=   r(t,S_t, V_t, \pi_t) D_t, \,  \quad D_0 =1,
}
where $ r:[0,T] \x \RR^d  \x\RR \x \RR^d  \rightarrow \RR$ is the interest rate, 
$S_t=\{S^i_t\}_{i=1}^d$ is the $d$-dimensional risky asset
price process, $V_t$ is the (real-valued) value process, % of hedger's trading strategy, 
and $\pi_t = \{\pi^i_t\}_{i=1}^d$ is the portfolio process
with $\pi^i_t$ being the number assets of the $i$th stock.  %We sssume that $d=m+l$.
The evolution of $S^i_t$ is assumed to be governed by the
SDE %\setlength{\parskip}{-3pt}
\aaa{
\lb{eqSDEpriceteugels}
dS^i_t = S^i_t \big\{\td f_i(t,S_t,V_t,\pi_t) dt +  \sum_{j=1}^d  \si_i^{\al_j}(t,S_{t-},V_{t-}) d M^{\al_j}_t\big\}
}
with the non-random initial condition $S^i_0>0$.  In \rf{eqSDEpriceteugels}, $\td f_i$ and $\si_i^{\al_j}$ are real-valued functions 
defined on spaces of appropriate dimensions. Further, for  $j=1,2, \ldots, m$,  $\al_j = j$ and $M^{\al_j}_t=B^j_t$, while 
for $j=m+1, \ldots, d$,  $\al_j$'s are arbitrarily picked
multiindexes from the set 
$\{(ik), i=1, \ldots, l, k=2,4,\ldots\}$ and $M^{\al_j}_t=H^{\al_j}_t$. Remark, that the index $k$ takes only even values.
%Furthermore, we assume that   the $l$ polynomials are chosen  s.t they  have a global minimum (e.g. pick even-degree polynomials.)

The value process $V_t$ represents the wealth
of a ``large'' investor who holds $d$ stocks and money on a deposit.  The investor is assumed large, so
the coefficients in our model would depend on $V_t$, $S_t$, and $\pi_t$.
%Let $\pi_i$ be the investment in the $i$th stock. and $V_t - \sum_{i=1}^d \pi_i$  is the investment in $D.$
%Now, borrowing from the case $d=1$ by \cite{chan} (p. 516),

We define an \textit{admissible hedging strategy}  
 as a pair of predictable processes $(\pi_t,\pi^0_t)$ such that $V_t = \sum_{i=1}^d \pi^i_t S^i_t + \pi^0_t D_t$  and $V_T=h(S_T)$,
 where $h(S_T)$ is the payoff at maturity $T$.
 Note that the solution of \rf{bond} takes the form $D_t = \exp\{\int_0^t r_s ds\}$, where $r_s =  r(s,S_s, V_s, \pi_s)$. Let $A_t =  \exp\{-\int_0^t r_s ds\}$.
 Define $\hat S^i_t=A_t S^i_t$ and  $\hat V_t=A_t V_t$ to  be the \textit{discounted stock price} and \textit{discounted value process}, respectively.
 Furthermore, we define the \textit{cumulative cost process}   $C_t= \hat V_t - \sum_{i=1}^d \int_0^t \pi^i_s d \hat S^i_s.$   
 %We say that the strategy $(\pi_t,\pi^0_t)$  is \textit{mean-self-financing} if it is admissible  and $C_t$ is a square-integrable martingale. 
 We say that the strategy is \textit{optimal}, if it is admissible  and $C_t$ is a square-integrable martingale
 strongly orthogonal to the martingale part of each $\hat S^i_t$.
\begin{Lem}
The representations $\hat V_t = \sum_{i=1}^d \int_0^t \pi^i_s d \hat S^i_s + C_t$ 
and  
\aaa{
\lb{repV}
V_t = V_0 + \sum_{i=1}^d \int_0^t \pi^i_s dS^i_s + \int_0^t \pi^0_s dD_s + \int_0^t D_sdC_s
}
are equivalent.
\end{Lem}
\begin{proof}
Since  $\<V, A\>_t= \<S^i, A \>_t= [V, A]_t=[ S^i, A]_t=0$, then by 
 It\^o's product formula, $d \hat S^i_t= A_t dS^i_t -A_t r_t S^i_t dt$ and
$d \hat V_t= A_t d  V_t  - r_t    A_t V_tdt$. Substituting these expressions 
%for $d \hat S^i_t$ and $d \hat V_t$ 
into the equation for $\hat V_t$ we obtain
$d V_t  = \sum_{i=1}^d \pi^i_t   d S^i_t +   ( V_t -  \sum_{i=1}^d \pi^i_t    S^i_t ) r_t dt+ D_tdC_t  = \sum_{i=1}^d \pi^i_t   d S^i_t + \pi^0_t  dD_t + D_tdC_t$.
\end{proof}
Now we derive a backward SDE (BSDE) for the process $V_t$ with representation \rf{repV}. First,
we substitute $dD_t$ and $dS^i_t$ with the right-hand sides of equations \rf{bond} and \rf{eqSDEpriceteugels}, respectively.
Since  $V_T = h(S_T)$, from \rf{repV} we obtain
\mmm{
\lb{V1}
V_t - \E[h(S_T)   - \int_t^T g(s,S_s, V_s, Z^{(\al)}_s) ds -  \sum_{j=1}^d \int_0^T Z^{\al_j}_s d M^{\al_j}_s - \int_0^T D_sdC_s| \mc F_t] \\
= \sum_{j=1}^d \int_0^t Z^{\al_j}_s d M^{\al_j}_s + \int_0^t D_sdC_s = \sum_{k=1}^l\sum_{j=1}^\infty \int_0^t   \hat Z^{(kj)}_s dH^{(kj)}_s,
}
where
 $g(t,s,v,z) = \td g(t,s,v,\mathbf s^{-1} \si(t,s,v)^{-1} z)$ with $\si(t,s,v)$ being the $d\x d$ matrix with the element $\si^{\al_j}_i$ 
 in the $j$th line and the $i$th column, 
$\td g(t,s,v,\pi)=\sum_{i=1}^d s_i\pi_i  f_i(t,s,v,\pi) +  (v- \sum_{i=1}^d s_i \pi_i)r(t,s,v,\pi)$, $\pi= (\pi_i)_{i=1}^d$, $s=(s_i)_{i=1}^d$,
and $\mathbf s = \diag\{s_1,\ldots, s_n\}$.
Further, $(\al)$ denotes the set of multiindexes $(\al) = \{ \al_1, \ldots, \al_d\}$  and  $Z^{(\al)}_t = (Z^{\al_1}_t,$ $ \ldots, Z^{\al_d}_t)$.
The last identity in \rf{V1}  follows from the predictable representation property of the system $\{H^{(kj)}_t\}_{j=1}^\infty$ for a fixed $k$
(with $\hat Z^{(kj)}_t$ being predictable processes)
and from the independence of the L\'evy processes $L^k_t$, $k=1,\ldots, l$. 
The BSDE for $V_t$ follows from \rf{V1}:
%Equation \rf{V1} impies the following BSDE for $V_t$
\mmm{ 
\lb{TeugelsproblemBSDE}
V_t =  h(S_T)   - \int_t^T g(s,S_s, V_s, Z^{(\al)}_s) ds   - \sum_{j=1}^d \int_t^T Z^{\al_j}_s d M^{\al_j}_s \\ -
 \sum_{(kj) \notin (\al)} \int_t^T   \hat Z^{(kj)}_s dH^{(kj)}_s.
}
Making the change of variable $\pi = \mathbf s^{-1}\si(t,s,v)^{-1} z$ and introducing the functions
 %$r(t,s,v,z) = \td r(t,s,v,\si(t,s,v)^{-1} z)$ and 
 $f_i(t,s,v,z) = \td f_i(t,s,v,  \mathbf s^{-1}\si(t,s,v)^{-1} z)$, we transform SDE \rf{eqSDEpriceteugels}
  to 
 \aaa{
 \lb{stock-bond}
% \begin{cases}
 dS^i_t = S^i_t \big\{f_i(t,S_t,V_t,Z^{(\al)}_t) dt +  \sum_{j=1}^d  \si_i^{\al_j}(t,S_{t},V_{t}) d M^{\al_j}_t\big\}.
 %dD_t=D_t \, r(t,S_t, V_t, Z^{(\al)}_t) dt, \quad D_0 =1.
 %\end{cases}
  }
  \begin{Lem} 
  \lb{lem8}
  FBSDEs \rf{TeugelsproblemBSDE}-\rf{stock-bond} are equivalent to 
  \eq{
  \lb{eqFBSDEsitu} 
S^i_t = S^i_0+ \int_0^t S^i_s \,  f_i(s,S_s, V_s,\mc Z_s, \mc{\hat Z}_s(\fdot)) ds + \int_0^t  S^i_s \, \si_i(s,S_s, V_s) \, dB_s  \\
 \hspace{1.4cm} +  \int_0^t \int_{\RR^l  }S_s^i \,\psi_i(s,S_{s-},V_{s-},u) \td N(ds,du), \qquad i =1,\ldots, d, \\
 V_t = h(S_T) -  \int_t^T   g (s,S_s, V_s,\mc Z_s, \mc{\hat Z}_s(\fdot)) ds -  \int_t^T \mc Z_s  dB_s  \\
   \hspace{2cm}   - \int_t^T  \int_{\RR^l  }\mc{\hat Z}_s(u)  \td N (ds,du),
}
where for $u= (u_1,\ldots, u_l)$, $\psi_i (t,s,v,u) = \sum_{q=m+1}^d   \si_i^{\al_q} (t,s,v) p_{\al_q}(u_{\al_q})$
with $u_{\al_q} = u_k$  if $\al_q = (kj)$.
Further, 
%the connection between the processes $(\mc Z_t, \mc {\hat Z}(t,\fdot))$ and $(Z^{(\al)}_t, \hat Z^{(kj)}_t)$ $(k=1,\ldots, l$,
%$j=1,2, \ldots)$ is the following:
  $\mc Z_t=(Z^{\al_1}_t, \ldots, Z^{\al_m}_t)$ and for each $k\in \{1,\ldots, l\}$, $\hat Z_t^{(kj)}$ are the 
  components of the decomposition of  
  %$\mc{\hat Z}_t(0, \ldots, \underbrace{u_k}_k, \ldots, 0)$
  $\mc{\hat Z}_t(0, \ldots, u_k, \ldots, 0)$
 with respect to the basis  of polynomials $p_{(kj)}(u_k)$ in the space $L_2(\nu^k(du_k))$, while  
 $(Z^{\al_{m+1}}_t, \ldots, Z^{\al_d}_t) = \{\hat Z_t^{(kj)}\}_{(kj)\in (\al)}$.
  Finally, $\td N$ is the compensated Poisson random measure for the L\'evy process $(L^1_t, \ldots, L^l_t)$.
  \end{Lem}
   \begin{Rmk} 
   \rm
With a slight abuse of notation, in the coefficients $f$ and $g$, we write 
$\mc Z_t$ instead of $(Z^{\al_1}_t, \ldots, Z^{\al_m}_t)$ and
$\mc {\hat Z}_t(\fdot)$ instead of $(Z^{\al_{m+1}}_t, \ldots, Z^{\al_d}_t)$. The dependence
on $\mc {\hat Z}_t(\fdot)$  is understood as the dependence on its $d-m$ components $(Z^{\al_{m+1}}_t, \ldots, Z^{\al_d}_t)$. 
%Further,  $\mc{ \hat Z}(t,u_k)$ is the restriction $\mc{ \hat Z}(t,u)|_{R_k}$ to the line $R_k$ generated by the basis vector $e_k$.
  \end{Rmk}
  \begin{proof}[Proof of Lemma \ref{lem8}]
  Note that for each $k$, the system $\{H^{(kj)}_t\}_{j=1}^\infty$ has
the predictable representation property. Therefore, 
  \mm{
  \int_t^T  \int_{\RR^l} \mc{\hat Z}_s(u)  \td N (ds,du) = \sum_{k=1}^l \int_t^T  \int_{R_k} \mc{\hat Z}_s(0, \ldots, u_k, \ldots, 0)  \td N^k (ds,du_k) \\
  =  \sum_{k=1}^l \sum_{j=1}^\infty \int_t^T   \hat Z^{(kj)}_s dH^{(kj)}_s,
  }
where $R_k=\{te_k, t\in \RR\}$ with $\{e_k\}_{k=1}^l$ being an orthonormal basis in $\RR^l$, and $\td N^k (t,\fdot)$ is the compensated
  Poisson random measure for $L^k_t$, which, by the independence of $L^k_t$'s, 
  is the restriction of $\td N(t,\fdot)$ to $R_k$. Since, by Lemma \ref{hi-rep}, $H^{(kj)}_t= \int_{R_k} p_{(kj)}(u_k) \td N^k(t, du_k)$,
  we obtain that  in $L_2(\nu^k)$, $\mc{\hat Z}_t(0,\ldots, u_k, \ldots, 0)  = \sum_{j=1}^\infty \hat Z^{(kj)}_t  p_{(kj)}(u_k)$ a.s.
  Moreover, for each $k$, 
  the system of polynomials $\{p_{(kj)}\}_{j=1}^\infty$
 is orthonormal in $L_2(\nu^k)$ by the orthonormality of $H^{(kj)}_t$'s. Finally, since for each $(kj)$, $p_{(kj)}(0) = 0$, we obtain
  \mm{
  \int_{\RR^l}  \sum_{q=m+1}^d \si_i^{\al_{q}} (t,x,y) p_{\al_q}(u_{\al_q})  \td N(dt,du) \\
  = \sum_{q=m+1}^d \si_i^{\al_q} (t,x,y)  \int_{R_{\al_q}}  p_{\al_q}(u_{\al_q}) \td N^{\al_k}(dt,du_{\al_q})
  = \hs \sum_{k=m+1}^d   \si_i^{\al_q} (t,x,y) dH^{\al_q}_t.
  }
  where $R_{\al_q} = R_k$, $\td N^{\al_q} = \td N^k$, and $u_{\al_q} = u_k$ for $\al_q = (kj)$.
  \end{proof}
 
\begin{Rmk} 
Since $\{\al_j\}_{j=m+1}^d$ are 
multiindexes picked from  the set $\{(ik), i=1, \ldots, l, k=2,4,\ldots\}$, then each polynomial $p_{\al_j}$ is  of even degree,
and, therefore, achieves a finite global minimum, which we denote by $A_j$.
 \end{Rmk}
 Assumption (A1)  below guarantees the existence and uniqueness of solution to FBSDEs \rf{TeugelsproblemBSDE}--\rf{stock-bond}. 
 \bi 
 \item[\bf (A1)] The coefficients of FBSDEs \rf{eqFBSDEsitu}  satisfy the hypotheses of  Theorem 3.1 in \cite{wu99} (p. 436).
 \ei
 \begin{Lem}
 \lb{lem10}
 Assume (A1). Then, there exists a unique $\mc F_t$-adapted solution $(S_t,V_t,Z^{(\al)}_t)$ to FBSDEs
\rf{TeugelsproblemBSDE}--\rf{stock-bond} such that $(S_t,V_t)$ has c\`adl\`ag paths and $Z^{(\al)}_t$ is predictable.
 \end{Lem}
 \begin{proof}
 Under (A1), Theorem 3.1 in \cite{wu99} guarantees the existence of a unique 
 $\F_t$-adapted solution  
 $(S_t, V_t, \mc Z_t, \mc {\hat Z}_t(\cdot))$ to FBSDEs \rf{eqFBSDEsitu} such that  $(S_t,V_t)$  is c\`adl\`ag
 and $(\mc Z_t, \mc {\hat Z}_t(\cdot))$ is predictable.
 By Lemma \ref{lem8}, this is equivalent to the existence of a unique $\F_t$-adapted solution  $(S_t,V_t,Z^{(\al)}_t)$ to
 \rf{TeugelsproblemBSDE}-\rf{stock-bond}.
 \end{proof}
 %To prove main result of our article, 
 %which is the existence of the optimal hedging strategy,
 %We will need 
 Assumptions (A2)--(A4) below guarantee the positivity of the prices $S^i_t$, the non-negativity
 of $V_t$, and the existence of the optimal strategy.
 \bi
  \item[\bf (A2)]  $\det\{\si(t,s,v)\} \neq 0$ for all $(t,s,v) \in [0,T] \x \RR^d \x \RR$.
  \item[\bf (A3)] For all $(t,s,v) \in [0,T] \x \RR^d \x \RR$, $i\in \{1,\ldots d\}$, and $j\in \{m+1,\ldots, d\}$, $\si^{\al_j}_i(t,s,v)>0$. Moreover,
   if $A_j<0$, then $\si^{\al_j}_i(t,x,v) |A_j|  < (d-m)^{-1}$.
 %where $A_{\al_j}$ is the (negative) global minimum of the polynomial $p_{\al_j}(x).$ %Furthermore, it holds  $\int_0^T \|\si(s,x,y)\|^2 ds \lt C.$ 
 \item[\bf (A4)]  If $(S_t,V_t,\mc Z_t, \mc {\hat Z}_t(\fdot))$ is the $\mc F_t$-adapted solution to FBSDEs \rf{eqFBSDEsitu}, then
 the random function $(\om, t,y,z,\hat z) \mto g(t,S_t,y,z,\hat z)$
 satisfies condition $(A_\gamma)$ in \cite{Royer} (p. 1362). Moreover, $h(S_T)\gt 0$ a.s.
 \ei
 The main result of our paper is the following.
 \begin{Thm}
Let (A1)--(A4) hold, and let $(S_t,V_t, Z^{(\al)}_t)$ be the solution to FBSDEs \rf{TeugelsproblemBSDE}-\rf{stock-bond}.
 Then,  $S^i_t>0$, $i=1,\ldots, d$, and $V_t\gt 0$ a.s. Moreover,
 the pair of stochastic processes $(\pi_t,\pi^0_t)$, where
$\pi_t = \diag\{S^1_t, \ldots, S^d_t\}^{-1}\sg^{-1}(t,S_t,V_t) Z^{(\al)}_t$ and  $\pi^0_t  = \hat V_t - \sum_{i=1}^d \pi^i_s \hat S^i_t$,
is the optimal hedging strategy.
\end{Thm}
 \begin{proof} 
 Note that the above representation for $\pi_t$ holds by  construction.
 %$\pi_t = \diag\{S^1_t, \ldots, S^d_t\}^{-1}\sg^{-1}(t,S_t,V_t) Z^{(\al)}_t$. 
%Let us prove the positivity of the prices $S^i_t$.
 Next, by the representation for the function $\psi_i$, obtained in Lemma \ref{lem8}, and by (A3),
% $\psi_i(t, S_{t-}, V_{t-}, \Dl L_t)= \sum_{j =m+1}^d p_{\al_j}(\Dl L_t^j) \si_i^{\al_j}(t,S_{t-},V_{t-})$.
 $\inf_{t>0} \psi_i(t, S_{t-}, V_{t-}, \Dl L_t) > -1$.  Therefore, $S^i_t$ can be represented by the Dol\'eans-Dade  exponential
 which is finite a.s.:
 \aa{
&S^i_t = S^i_0\,  e^{\int_0^t \big(\td f_i(s, \, S_s,V _s, \pi_s) -  \frac{\| \si_i(s,\, S_s,V_s) \|^2}{2}\big) ds  +  \int_0^t  \si_i (s,\, S_s,V_s)  dB_s
+ \int_0^t \int_{\RR^l} \psi_i (s,\, S_{s-},V_{s-},u)  \td N(ds,du)}  \\
& \x
 \prod_{0 \lt s \lt t}\big(1+ \psi_i(s, S_{s-}, V_{s-},\Dl L_s)\big) e^{ - \psi_i(s, \, S_{s-}, V_{s-},\, \Dl L_s)},
 }
 where $\sg_i = (\sg^{\al_j}_i)_{j=1}^d$.
Therefore,  for all $i$, $S^i_t>0$ a.s. Let us prove the non-negativity of $V_t$. To this end, we apply the comparison theorem from 
\cite{Royer} (Theorem 2.5, p. 1362)
 to the BSDE in \rf{eqFBSDEsitu}, considered with respect to $(V_t, \mc Z_t, \mc {\hat Z}_t(\fdot))$, whereas the process $S_t$ is
 fixed and assumed known from Lemma \ref{lem10}. Note that, by the  definition, $g(t,S_t,0,0,0) = 0$. Therefore, 
 we compare the solution $(V_t, \mc Z_t,  \mc {\hat Z}_t(\fdot))$ with the identically zero solution to the BSDE whose generator is 
 the same as in \rf{eqFBSDEsitu} but
 the final condition is zero. Remark that (A4) implies the assumptions of the comparison theorem in \cite{Royer}. 
 Thus, by Theorem 2.5 in \cite{Royer}, $V_t \gt 0$ for all $t$ a.s. 
 
Note that, by \rf{V1}, $C_t = V_0 + \sum_{(kj)\notin(\al)} \int_0^t A_s \hat Z^{(kj)}_s dH^{(kj)}_s$, and, therefore, it is 
a square integrable martingale. Moreover, $C_t$ is (weakly) orthogonal to the stable subspace $\mc S$ generated by $\{M^{\al_j}_t\}_{j=1}^d$,
which follows from Theorem 35 of \cite{Protter} (p. \hsp 149) and from the strong orthogonality of the martingales $M^{\al_j}_t$.
By Theorem 36 of \cite{Protter} (p. \hsp 150), $C_t$ is strongly orthogonal to $\mc S$. It remains to note that the martingale parts of 
$\{\hat S^i_t\}_{i=1}^d$  belong to $\mc S$. %The theorem is proved.
\end{proof}
\begin{Cor} 
The F\"ollmer-Schweizer decomposition of the discounted contingent claim $A_T h(S_T)$ takes the form
\aa{
A_T h(S_T)= V_0 + \sum_{i=1}^d \int_0^T \pi^i_t \, d\hat S^i_t +\sum_{(kj)\notin(\al)} \int_0^T A_t \hat Z^{(kj)}_t dH^{(kj)}_t.
}
\end{Cor}

%\section*{References}

\end{document}